\def\year{2022}\relax
\documentclass[letterpaper]{article} 
\usepackage{aaai22}  
\usepackage{times}  
\usepackage{helvet}  
\usepackage{courier}  
\usepackage[hyphens]{url}  
\usepackage{graphicx} 
\usepackage{natbib}  
\usepackage{caption} 
\DeclareCaptionStyle{ruled}{labelfont=normalfont,labelsep=colon,strut=off} 
\usepackage{amssymb}
\usepackage{amsmath}
\usepackage{amsthm}
\usepackage{mathtools}
\DeclarePairedDelimiter\abs{\lvert}{\rvert}
\usepackage{booktabs}
\newtheorem{proposition}{Proposition}
\usepackage{comment}
\usepackage{subfigure}
\usepackage{mwe}
\usepackage[table]{xcolor}
\usepackage{algorithm}
\usepackage{algorithmic}
\usepackage{multirow}
\usepackage{newfloat}
\usepackage{listings}
\floatstyle{ruled}
\newfloat{listing}{tb}{lst}{}
\floatname{listing}{Listing}
\title{Obtaining Calibrated Probabilities with Personalized Ranking Models}
\author{
    Wonbin Kweon, SeongKu Kang, Hwanjo Yu\thanks{Corresponding author}
}
\affiliations{
    Pohang University of Science and Technology, South Korea\\

    \{kwb4453, seongku, hwanjoyu\}@postech.ac.kr
}

\begin{document}
\maketitle

\begin{abstract}
For personalized ranking models, the well-calibrated probability of an item being preferred by a user has great practical value.
While existing work shows promising results in image classification, probability calibration has not been much explored for personalized ranking.
In this paper, we aim to estimate the calibrated probability of how likely a user will prefer an item. 
We investigate various parametric distributions and propose two parametric calibration methods, namely Gaussian calibration and Gamma calibration.
Each proposed method can be seen as a post-processing function that maps the ranking scores of pre-trained models to well-calibrated preference probabilities, without affecting the recommendation performance.
We also design the unbiased empirical risk minimization framework that guides the calibration methods to the learning of true preference probability from the biased user-item interaction dataset.
Extensive evaluations with various personalized ranking models on real-world datasets show that both the proposed calibration methods and the unbiased empirical risk minimization significantly improve the calibration performance.
\end{abstract}

\section{Introduction}
\noindent Personalized ranking models aim to learn the ranking scores of items, so as to produce a ranked list of them for the recommendation \cite{bpr09}.
However, their prediction results provide an incomplete estimation of the user's potential preference for each item; the semantic of the same ranking position differs for each user.
One user might like his third item with the probability of 30\%, whereas the other user likes her third item with 90\%.
Accurately estimating the \textit{probability} of an item being preferred by a user has great practical value \cite{iso12}.
The preference probability can help the user choose the items with high potential preference and the system can raise user satisfaction by pruning the ranked list by filtering out items with low confidence \cite{pruning09}.
To ensure reliability, the predicted probabilities need to be \textit{calibrated} so that they can accurately indicate their ground truth correctness likelihood.
In this paper, our goal is to obtain the well-calibrated probability of an item matching a user's preference based on the ranking score of the pre-trained model, without affecting the ranking performance.

While recent methods \cite{cal17, kull2019beyond, rahimi2020intra} have successfully achieved model calibration for image classification, it has remained a long-standing problem for personalized ranking.
A pioneering work \cite{iso12} firstly proposed to predict calibrated probabilities from the scores of pre-trained ranking models by using isotonic regression \cite{iso72}, which is a simple non-parametric method that fits a monotonically increasing function.
Although it has shown some effectiveness, there is no subsequent study about \textit{parametric} calibration methods in the field of personalized ranking despite their richer expressiveness than non-parametric methods.

In this paper, we investigate various parametric distributions, and from which we propose two calibration methods that can best model the score distributions of the ranking models.
First, we define three desiderata that a calibration function for ranking models should meet, and show that existing calibration methods have the insufficient capability to model the diverse populations of the ranking score.
We then propose two parametric methods, namely Gaussian calibration and Gamma calibration, that satisfy all the desiderata.
We demonstrate that the proposed methods have a larger expressive power in terms of the parametric family and also effectively handles the imbalanced nature of ranking score populations compared to the existing methods \cite{platt99, cal17}.
Our methods are post-processing functions with \textit{three} learnable parameters that map the ranking scores of pre-trained models to calibrated posterior probabilities.

To optimize the parameters of the calibration functions, we can use the log-loss on the held-out validation sets \cite{cal17}.
The challenge here is that the user-item interaction datasets are implicit and missing-not-at-random \cite{ips16, saito19}.
For each user-item pair, the label is 1 if the interaction is observed, 0 otherwise.
An unobserved interaction, however, does not necessarily mean a negative preference, but the item might have not been exposed to the user yet.
Therefore, if we fit the calibration function with the log-loss computed naively on the implicit datasets, the mapped probabilities may indicate biased likelihoods of users' preference on items.
To tackle this problem, we design an unbiased empirical risk minimization framework by adopting Inverse Propensity Scoring \cite{ips94}.
We first decompose the interaction variable into two variables for observation and preference, and adopt an inverse propensity-scored log-loss that guides the calibration functions toward the true preference probability.

Extensive evaluations with various personalized ranking models on real-world datasets show that the proposed calibration methods produce more accurate probabilities than existing methods in terms of calibration measures like ECE, MCE, and NLL.
Our unbiased empirical risk minimization framework successfully estimates the ideal empirical risk, leading to performance gain over the naive log-loss.
Furthermore, reliability diagrams show that Gaussian calibration and Gamma calibration predict well-calibrated probabilities across all probability range.
Lastly, we provide an in-depth analysis that supports the superiority of the proposed methods over the existing methods.

\section{Preliminary \& Related Work}
\subsection{Personalized Ranking}
Let $\mathcal{U}$ and $\mathcal{I}$ denote the user space and the item space, respectively.
For each user-item pair $(u,i)$ of $u \in \mathcal{U}$ and $i \in \mathcal{I}$, a label $Y_{u,i}$ is given as 1 if their interaction is observed and 0 otherwise.
It is worth noting that unobserved interaction ($Y_{u,i}=0$) may indicate the negative preference or the unawareness, or both.
A personalized ranking model $f_\theta: \mathcal{U} \times \mathcal{I} \rightarrow \mathbb{R}$ learns the ranking scores of user-item pairs to produce a ranked list of items for each user.
$f_{\theta}$ is mostly trained with pairwise loss that makes the model put a higher score on the observed pair than the unobserved pair:
\begin{equation}
    \mathcal{L}_{pair} = \sum_{u \in \mathcal{U}, i,j \in \mathcal{I}} \ell(f_\theta(u,i), f_\theta(u,j)) Y_{u,i} (1-Y_{u,j}),
\end{equation}
where $\ell(\cdot, \cdot)$ is some convex loss function such as BPR loss \cite{bpr09} or Margin Ranking loss \cite{margin07}.
Note that the ranking score $f_\theta(u,i) \in \mathbb{R}$ is not bounded in $[0,1]$ and therefore cannot be used as a probability.

\subsection{Calibrated Probability}
To estimate $P(Y_{u,i}=1 | f_\theta(u,i))$, which is the probability of item $i$ being interacted with user $u$ given the pre-trained ranking score, we need a post-processing calibration function $g_{\phi}(s)$ that maps the ranking score $s=f_{\theta}(u,i)$ to the calibrated probability $p$.
Here, the calibration function for the personalized ranking has to meet the following \textbf{desiderata}:
(1) the function $g_\phi: \mathbb{R} \rightarrow [0,1]$ needs to take an input from the unbounded range of the ranking score to output a probability;
(2) the function should be \textit{monotonically increasing} so that the item with a higher ranking score gets a higher preference probability;
(3) the function needs enough expressiveness to represent diverse score distributions.

We say the probability $p$ is well-calibrated if it indicates the ground-truth correctness likelihood \cite{beta17}:
\begin{equation}
    \mathbb{E} [ Y | g_\phi(s) = p ] = p, \;\;\; \forall p \in [0,1].
\end{equation}
\begin{equation}
    g_\phi(f_\theta(u,i)) = p 
\end{equation}
For example, if we have 100 predictions with $p=0.3$, we expect 30 of them to indeed have $Y=1$ when the probabilities are calibrated.
Using this definition, we can measure the miscalibration of a model with Expected Calibration Error (ECE) \cite{bbq15}:
\begin{equation}
    \text{ECE}(g_\phi) = \mathbb{E} \big[ \abs{\mathbb{E}[Y|g_\phi(s)=p] - p} \big].
\end{equation}
However, since we only have finite samples, we cannot directly compute ECE with Eq.3.
Instead, we partition the [0,1] range of $p$ into $M$ equi-spaced bins and aggregate the value of each bin:
\begin{equation}
    \text{ECE}_M(g_\phi) = \sum_{m=1}^M \frac{\abs{B_m}}{N} \left| \frac{\sum_{k \in B_m} Y_k}{\abs{B_m}} - \frac{\sum_{k \in B_m} p_k}{\abs{B_m}} \right|,
\end{equation}
where $B_m$ is $m$-th bin and $N$ is the number of samples.
The first term in the absolute value symbols denotes the ground-truth proportion of positive samples (accuracy) in $B_m$ and the second term denotes the average calibrated probability (confidence) of $B_m$.
Similarly, Maximum Calibration Error (MCE) is defined as follows:
\begin{equation}
    \text{MCE}_M(g_\phi) = \max_{m \in \{1,..,M\}} \left| \frac{\sum_{k\in B_m} Y_k}{\abs{B_m}} - \frac{\sum_{k \in B_m} p_k}{\abs{B_m}} \right|.
\end{equation}
MCE measures the worst-case discrepancy between the accuracy and the confidence.
Besides the above calibration measures, Negative Log-Likelihood (NLL) also can be used as a calibration measure \cite{cal17}.

\subsection{Calibration Method}
Existing methods for model calibration are categorized into two groups: non-parametric and parametric methods.
Non-parametric methods mostly adopt the binning scheme introduced by the histogram binning \cite{hist01}.
The histogram binning divides the uncalibrated model outputs into $B$ equi-spaced bins and samples in each bin take the proportion of positive samples in the bin as the calibrated probability.
Subsequently, isotonic regression \cite{iso12} adjusts the number of bins and their width, Bayesian binning into quantiles (BBQ) \cite{bbq15} takes an average of different binning models for the better generalization.
In the perspective of our desiderata, however, none of them meets all three conditions (please refer to Appendix A).

The parametric methods try to fit calibration functions that map the output scores to the calibrated probabilities.
Temperature scaling \cite{cal17}, a well-known technique for calibrating deep neural networks, is a simplified version of Platt scaling \cite{platt99} that adopts Gaussian distributions with the same variance for the positive and the negative classes.
Beta calibration \cite{beta17} utilizes Beta distribution for the binary classification and Dirichlet calibration \cite{kull2019beyond} generalizes it for the multi-class classification.
While recent work \cite{rahimi2020intra, mukhoti2020calibrating} is focusing on parametric methods and shows promising results for image classification, they cannot be directly adopted for the personalized ranking.
Also, the above parametric methods do not satisfy all the desiderata (please refer to Appendix A).
In this paper, we propose two parametric calibration methods that satisfy all the desiderata for the personalized ranking models.

\section{Proposed Calibration Method}
\subsection{Revisiting Platt Scaling}
Platt scaling \cite{platt99} is widely used parametric calibration method, which is a generalized form of the temperature scaling \cite{cal17}:
\begin{equation}
    g_\phi^{\text{Platt}}(s) = \sigma(bs + c),
\end{equation}
where $\phi = \{b,c\}$ are learnable parameters and $\sigma(x) = 1/(1+\text{exp}(-x))$ is the sigmoid function.
In this section, we show that Platt scaling can be derived from the assumption that the class-conditional scores follow Gaussian distributions with the same variance.

We first set the class-conditional score distribution for the positive and the negative classes:
\begin{equation}
\begin{split}
    & p(s|Y=0) = (\sqrt{2\pi}\sigma_0)^{-1} \text{exp} [ - (s-\mu_0)^2 / 2\sigma_0^2 ], \\
    & p(s|Y=1) = (\sqrt{2\pi}\sigma_1)^{-1} \text{exp} [ - (s-\mu_1)^2 / 2\sigma_1^2 ], \\
\end{split}
\end{equation}
where $\mu_0, \mu_1 \in \mathbb{R}$, $\sigma_0^2, \sigma_1^2 \in \mathbb{R}^+$ are the mean and the variance of each Gaussian distribution.
Then, the posterior is computed as follows:
\begin{equation}
\begin{split}
    P(Y=1|s) & = \frac{\pi_1 p(s|Y=1)}{\pi_1 p(s|Y=1) + \pi_0 p(s|Y=0)} \\
    & = \frac{1}{1 + \pi_0 p(s|Y=0) / \pi_1 p(s|Y=1)} \\
    & = \frac{1}{1 + \text{exp}\big[ (\frac{1}{2\sigma_1^2} - \frac{1}{2\sigma_0^2})s^2 + (\frac{\mu_0}{\sigma_0^2}-\frac{\mu_1}{\sigma_1^2})s - c \big]} \\
    & = \sigma(as^2 + bs + c),
\end{split}
\end{equation}
where $\pi_0$ and $\pi_1$ are the prior probability for each class, $a = (2\sigma_0^2)^{-1} - (2\sigma_1^2)^{-1}$, $b = \mu_1/\sigma_1^2 - \mu_0/\sigma_0^2$, and $c=\mu_1^2 / (2\sigma_1^2) - \mu_0^2 / (2\sigma_0^2) + \text{log}(\pi_0 \sigma_1) - \text{log}(\pi_1 \sigma_0) \in \mathbb{R}$.
We can see that Platt scaling is a special case of Eq.8 with the assumption $a=0$ (i.e., the same variance for both class-conditional score distributions).

\subsection{Gaussian Calibration}
For personalized ranking, however, the usage of the same variance for both class-conditional score distributions is not desirable, because a severe imbalance between the two classes exists in user-item interaction datasets.
Since users have distinct preferences for item categories, preferred items take only a small portion ($\sim$10\% in real-world datasets) of the entire itemset.
Therefore, the score distribution of diverse unpreferred items and that of distinct preferred items are likely to have disparate variances.

To tackle this problem, we let the variance of each class-conditional score distribution be optimized with datasets, without any naive assumption of the same variance for both classes:
\begin{equation}
    g_\phi^{\text{Gaussian}}(s) = \sigma(as^2 + bs + c),
\end{equation}
where $\phi = \{a, b,c\}$ are learnable parameters and can be any real numbers.
Since $a=(2\sigma_0^2)^{-1} - (2\sigma_1^2)^{-1}$ can capture the different deviations of two classes during the training, we can handle the distinct distribution of each class.

\subsection{Gamma Calibration}
Gamma distribution is also widely adopted to model the score distribution of ranking models \cite{gamma99}.
Unlike Gaussian distribution that is symmetric about its mean, Gamma distribution can capture the skewed population of ranking scores that might exist in the datasets.
In this section, we set the class-conditional score distribution to Gamma distribution:
\begin{equation}
\begin{split}
    & p(s|Y=0) = \Gamma(\alpha_0)^{-1} \beta_0^{\alpha_0} s^{\alpha_0-1} \text{exp}(-\beta_0s), \\
    & p(s|Y=1) = \Gamma(\alpha_1)^{-1} \beta_1^{\alpha_1} s^{\alpha_1-1} \text{exp}(-\beta_1s), \\
\end{split}
\end{equation}
where $\Gamma(\cdot)$ is the Gamma function, $\alpha_0 , \alpha_1 , \beta_0, \beta_1 \in \mathbb{R}^+$ are the shape and the rate parameters of each Gamma distribution.
Then, the posterior is computed as follows:
\begin{equation}
\begin{split}
    P(Y=1|s) & = \frac{1}{1 + \pi_0 p(s|Y=0) / \pi_1 p(s|Y=1)} \\
    & = \frac{1}{1 + \frac{\pi_0 \beta_0^{\alpha_0} \Gamma(\alpha_1)}{\pi_1 \beta_1^{\alpha_1} \Gamma(\alpha_0)} s^{\alpha_0 - \alpha_1} \text{exp}[(\beta_1-\beta_0)s]} \\
    & = \frac{1}{1 + \text{exp}\big[ (\alpha_0-\alpha_1)\text{log}s + (\beta_1-\beta_0)s - c \big]} \\
    & = \sigma(a\text{log}s + bs + c),
\end{split}
\end{equation}
where $a = \alpha_1-\alpha_0$, $b = \beta_0-\beta_1$, and $c=\text{log}(\pi_1\beta_1^{\alpha_1}\Gamma(\alpha_0) / \pi_0\beta_0^{\alpha_0}\Gamma(\alpha_1)) \in \mathbb{R}$.
Therefore, Gamma calibration can be formalized as follows:
\begin{equation}
    g_\phi^{\text{Gamma}}(s) = \sigma(a\text{log}s + bs + c),
\end{equation}
where $\phi = \{a, b,c\}$ are learnable parameters.
Since Gamma distribution is defined only for the positive real number, we need to shift the score to make all the inputs positive: $s \leftarrow s-s_{\text{min}}$, where $s_{\text{min}}$ is the minimum ranking score. 

\subsection{Other Distributions}
Besides adopting Gaussian distribution or Gamma distribution for both classes, there have been proposed other parametric distributions for modeling the ranking scores.
Swets adopts two Exponential distributions \cite{exp69}, Manmatha proposes Gaussian distribution for the positive class and Exponential distribution for the negative class \cite{normexp01}, and Kanoulas proposes Gaussian distribution for the positive class and Gamma distribution for the negative class \cite{gammagaussian10}.
We also investigated these distributions, however, they either have the same form as the proposed calibration function or their posterior cannot satisfy our desiderata.
Please refer to Appendix B for more information.

\subsection{Monotonicity for Proposed Desiderata}
The proposed calibration methods naturally satisfy the first and the third of our desiderata:
(1) the proposed methods take the unbounded ranking scores and produce calibrated probabilities;
(2) the proposed methods have richer expressiveness than Platt scaling or temperature scaling, since they have a larger capacity in terms of the parametric family.
The last condition that our calibration methods need to meet is that they should be monotonically increasing for maintaining the ranking order.
To this end, we need linear constraints on the parameters of each method: $2as+b > 0$ for Gaussian calibration and $a/s+b>0$ for Gamma calibration (derivation of these constraints can be found in Appendix C).
Since these constraints are linear and we have only three learnable parameters, the optimization of constrained logistic regression is easily done in at most a few minutes with the existing module of Scipy \cite{scikit-learn}.

\section{Unbiased Parameter Fitting}
\subsection{Naive Log-loss}
After we formalize Gaussian Calibration and Gamma Calibration, we need to optimize their learnable parameters $\phi$.
A well-known way to fit them is to use log-loss on the held-out validation set, which can be the same set used for the hyperparameter tuning \cite{cal17, beta17}.
Since we only observe the interaction indicator $Y_{u,i}$, the naive negative log-likelihood is computed for a user-item pair as follows:
\begin{equation}
      \mathcal{L}_{\text{naive}} = - Y_{u,i} \, \text{log}( g_{\phi}(s_{u,i})) - (1-Y_{u,i}) \text{log}(1 - g_{\phi}(s_{u,i})).
\end{equation}
where $s_{u,i} = f_{\theta}(u,i)$ is the ranking score for the user-item pair.
Note that during the fitting of the calibration function $g_{\phi}(s)$, the parameters of the pre-trained ranking model $f_{\theta}(u,i)$ are fixed.

\subsection{Ideal Log-loss for Preference Estimation}
The observed interaction label $Y_{u,i}$, however, indicates the presence of user-item interaction, not the user's preference on the item.
Therefore, $Y_{u,i}=0$ does not necessarily mean the user's negative preference, but it can be that the user is not aware of the item.
If we fit the calibration function with $\mathcal{L}_{\textup{naive}}$, mapped probabilities could be biased towards the negative preference by treating the unobserved positive pair as the negative pair.
To handle this implicit interaction process, we borrow the idea of decomposing the interaction variable $Y_{u,i}$ into two independent binary variables \cite{ips16}: 
\begin{equation}
\begin{split}
    Y_{u,i} & = O_{u,i} \cdot R_{u,i}, \\
    P(Y_{u,i}=1) & = P(O_{u,i}=1) \cdot P(R_{u,i}=1) \\
    & = \omega_{u,i} \cdot \rho_{u,i},
\end{split}
\end{equation}
where $O_{u,i}$ is a binary random variable representing whether the item $i$ is observed by user $u$, and $R_{u,i}$ is a binary random variable representing whether the item $i$ is preferred by user $u$.
The user-item pair interacts ($Y_{u,i}=1$) when the item is observed ($O_{u,i}=1$) and preferred ($R_{u,i}=1$) by the user.

The goal of this paper is to estimate the probability of an item being \textit{preferred} by a user, not the probability of an item being \textit{interacted} by a user.
Therefore, we need to train $g_{\phi}(s)$ for predicting $P(R=1|s)$ instead of $P(Y=1|s)$\footnote{We can replace $Y_{u,i}$ with $R_{u,i}$ in Eq.2$\sim$12.}.
To this end, we need a new ideal loss function that can guide the optimization towards the true preference probability:
\begin{equation}
      \mathcal{L}_{\text{ideal}} = - R_{u,i} \, \text{log}(g_{\phi}(s_{u,i})) - (1-R_{u,i}) \text{log}(1 - g_{\phi}(s_{u,i})).
\end{equation}
The ideal loss function enables the calibration function to learn the unbiased preference probability.
However, since we cannot observe the variable $R_{u,i}$ from the training set, the ideal log-loss cannot be computed directly.

\subsection{Unbiased Empirical Risk Minimization}
In this section, we design an unbiased empirical risk minimization (UERM) framework to obtain the ideal empirical risk minimizer.
We deploy the Inverse Propensity Scoring (IPS) estimator \cite{ips94}, which is a technique for estimating the counterfactual outcome of a subject under a particular treatment.
The IPS estimator is widely adopted for the unbiased rating prediction \cite{ips16, ips19} and the unbiased pairwise ranking \cite{joachims2017unbiased, saito19}.
For a user-item pair, the inverse propensity-scored log-loss for the unbiased empirical risk minimization is defined as follows:
\begin{equation}
      \mathcal{L}_{\text{UERM}} = - \frac{Y_{u,i}}{\omega_{u,i}} \, \text{log}( g_{\phi}(s_{u,i})) - (1-\frac{Y_{u,i}}{\omega_{u,i}}) \text{log}(1 - g_{\phi}(s_{u,i})),
\end{equation}
where $\omega_{u,i} = P(O_{u,i}=1)$ is called \textit{propensity score}. 
\begin{proposition}
$\hat{\mathcal{R}}_{\textup{UERM}}(g_{\phi}|\omega)$, which is the empirical risk of $\mathcal{L}_{\textup{UERM}}$ on validation set with true propensity score $\omega$, is equal to $\hat{\mathcal{R}}_{\textup{ideal}}(g_{\phi})$, which is the ideal empirical risk.
\end{proposition}
\noindent The proof can be found in Appendix D.
This proposition shows that we can get the unbiased empirical risk minimizer by $\phi^{\textup{UERM}} = \textup{argmin}_{\phi}\{ \hat{\mathcal{R}}_{\text{UERM}}(g_{\phi}|\omega)\}$ when only $Y_{u,i}$ is observed.

The remaining challenge is to estimate the propensity score $\omega_{u,i}$ from the dataset.
There have been proposed several techniques for estimating the propensity score such as Naive Bayes \cite{ips16} or logistic regression \cite{rosenbaum2002overt}.
However, the Naive Bayes needs unbiased held-out data for the missing-at-random condition and the logistic regression needs additional information like user demographics and item categories.
In this paper, we adopt a simple way that utilizes the popularity of items as done in \cite{saito19}:
$\hat{\omega}_{u,i} = ( \sum_{u \in \mathcal{U}} Y_{u,i}/\text{max}_{i \in \mathcal{I}} \sum_{u \in \mathcal{U}} Y_{u,i} )^{0.5}.$
While one can concern that this estimate of propensity score may be inaccurate, Schnabel \cite{ips16} shows that we merely need to estimate better than the naive uniform assumption.
We provide an experimental result that demonstrates our estimate of the propensity score shows comparable performance with Naive Bayes and Logistic Regression that use additional information (Appendix F).

For deeper insights into the variability of the estimated empirical risk, we investigate the bias when the propensity scores are inaccurately estimated.
\begin{proposition}
The bias of $\, \hat{\mathcal{R}}_{\textup{UERM}}(g_{\phi}|\hat{\omega})$ induced by the inaccurately estimated propensity scores $\hat{\omega}$ is  $\frac{1}{\abs{\mathcal{D}_{\textup{val}}}} \sum_{(u,i) \in \mathcal{D}_{\textup{val}}} \rho_{u,i} \left(\frac{\omega_{u,i}}{\hat{\omega}_{u,i}}-1 \right) \textup{log}\left( \frac{g_{\phi}(s_{u,i})}{1-g_{\phi}(s_{u,i})} \right)$.
\end{proposition}
\noindent The proof can be found in Appendix D.
Obviously, the bias is zero when the propensity score is correctly estimated.
Furthermore, we can see that the magnitude of the bias is affected by the inverse of the estimated propensity score.
This finding is consistent with the previous work \cite{clip19} that proposes to adopt a propensity clipping technique to reduce the variability of the bias.
In this work, we use a simple clipping technique $\hat{\omega}_{u,i} \leftarrow \max\{\hat{\omega}_{u,i}, 0.1\}$ that can prevent the item with extremely low popularity from amplifying the bias \cite{saito19}.

\begin{table*}[t]
    \centering\fontsize{9}{10}\selectfont
    \begin{tabular}{cc|ccccc|ccccc}
        \toprule
         & & \multicolumn{5}{c|}{Yahoo!R3} & \multicolumn{5}{c}{Coat} \\
        \toprule
        Type & Methods & BPR & NCF & CML & UBPR & LGCN & BPR & NCF & CML & UBPR & LGCN \\
        \midrule
        \multirow{2}{*}{uncalibrated} & MinMax & 0.4929 & 0.4190 & 0.3152 & 0.3004 & 0.2258 & 0.1790 & 0.4624 & 0.1834 & 0.1920 & 0.2350\\
         & Sigmoid & 0.3065 & 0.0729 & 0.0526 & 0.2516 & 0.3024 & 0.2196 & 0.1422 & 0.0647 & 0.1415 & 0.0508\\
        \midrule
        \multirow{3}{*}{non-parametric} & Hist & 0.0161 & 0.0133 & 0.0641 & 0.0130 & 0.0194 & 0.0552 & 0.0230 & 0.0161 & 0.0514 & 0.0470\\
         & Isotonic & 0.0146 & 0.0130 & 0.0635 & 0.0127 & 0.0154 & 0.0474 & 0.0159 & 0.0160 & 0.0490 & 0.0453\\
         & BBQ & 0.0136 & 0.0137 & 0.0634 & 0.0140 & 0.0165 & 0.0552 & 0.0178 & 0.0198 & 0.0459 & 0.0494\\
        \midrule
        \multirow{4}{*}{\shortstack{parametric \\ w/ $\mathcal{L}_{\textup{naive}}$}} & Platt & 0.0126 & 0.0146 & 0.0515 & 0.0107 & 0.0099 & 0.0441 & 0.0245 & 0.0203 & 0.0423 & 0.0407\\
         & Beta & 0.0127 & 0.0144 & 0.0504 & 0.0105 & 0.0150 & 0.0451 & 0.0258 & 0.0270 & 0.0416 & 0.0407\\
         & Gaussian & 0.0129 & 0.0104 & 0.0486 & 0.0105 & 0.0073 & 0.0436 & 0.0264 & 0.0245 & 0.0410 & 0.0404\\
         & Gamma & 0.0108 & 0.0145 & 0.0512 & 0.0107 & 0.0098 & 0.0424 & 0.0239 & 0.0208 & 0.0405 & 0.0406\\
        \midrule
          & Platt & 0.0106 & 0.0129 & 0.0303 & 0.0100 & 0.0070 & 0.0411 & 0.0120 & 0.0155 & 0.0354 & 0.0224\\
        parametric & Beta & 0.0109 & 0.0132 & 0.0305 & 0.0094 & 0.0076 & 0.0414 & 0.0075 & 0.0183 & 0.0375 & 0.0266\\
        w/ $\mathcal{L}_{\textup{UERM}}$ & \cellcolor{gray!20}Gaussian & \cellcolor{gray!20}0.0106 & \cellcolor{gray!20}\textbf{0.0096} & \cellcolor{gray!20}\textbf{0.0285} & \cellcolor{gray!20}\textbf{0.0070} & \cellcolor{gray!20}\textbf{0.0061} & \cellcolor{gray!20}0.0393 & \cellcolor{gray!20}0.0062 & \cellcolor{gray!20}\textbf{0.0147} & \cellcolor{gray!20}\textbf{0.0323} & \cellcolor{gray!20}\textbf{0.0208}\\
         & \cellcolor{gray!20}Gamma & \cellcolor{gray!20}\textbf{0.0100} & \cellcolor{gray!20}0.0117 & \cellcolor{gray!20}0.0287 & \cellcolor{gray!20}0.0085 &\cellcolor{gray!20}0.0065 & \cellcolor{gray!20}\textbf{0.0390} & \cellcolor{gray!20}\textbf{0.0061} &\cellcolor{gray!20}0.0148 &\cellcolor{gray!20} 0.0326 & \cellcolor{gray!20}0.0215\\
        \midrule
        & \textit{Improv} & 5.85\% & 25.35\% & 5.94\% & 25.85\% & 12.86\% & 5.21\% & 18.67\% & 5.41\% & 8.81\% & 7.14\% \\
        \bottomrule
    \end{tabular}
    \caption{Expected Calibration Error of each calibration method applied on five personalized ranking models. Numbers in boldface are the best results and \textit{Improv} denotes the improvement of the best proposed method over the best competitor (Platt or Beta with $\mathcal{L}_{\text{UERM}}$).}
    \label{tab:main}
\end{table*}

\begin{figure*}[h!]
\centering 
\subfigure{\includegraphics[width=0.18\linewidth]{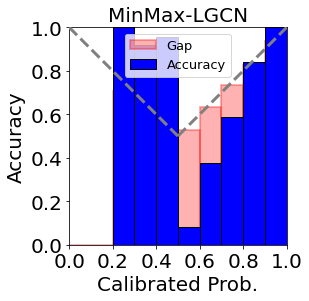}}
\subfigure{\includegraphics[width=0.18\linewidth]{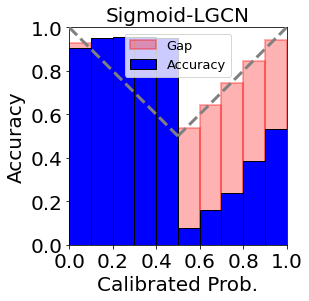}}
\subfigure{\includegraphics[width=0.18\linewidth]{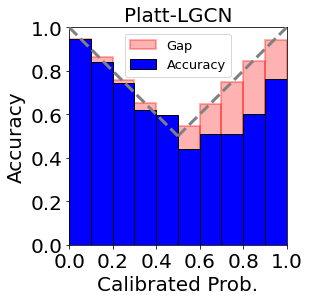}}
\subfigure{\includegraphics[width=0.18\linewidth]{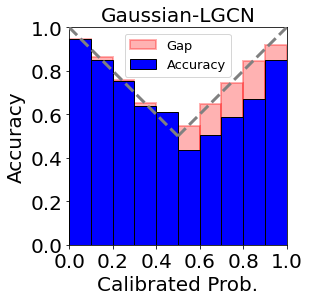}}
\subfigure{\includegraphics[width=0.18\linewidth]{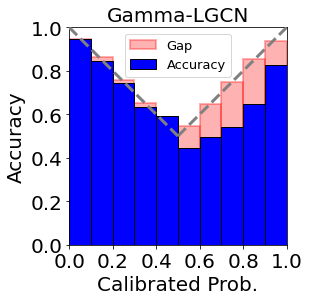}}
\subfigure{\includegraphics[width=0.18\linewidth]{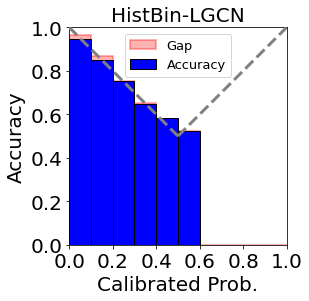}}
\subfigure{\includegraphics[width=0.18\linewidth]{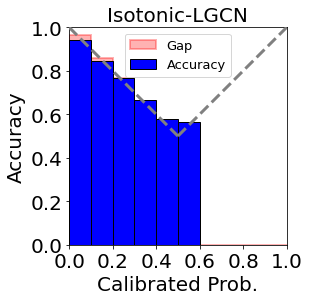}}
\subfigure{\includegraphics[width=0.18\linewidth]{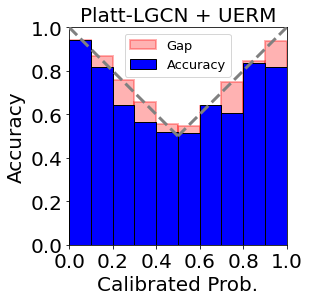}}
\subfigure{\includegraphics[width=0.18\linewidth]{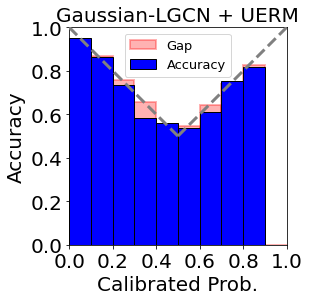}}
\subfigure{\includegraphics[width=0.18\linewidth]{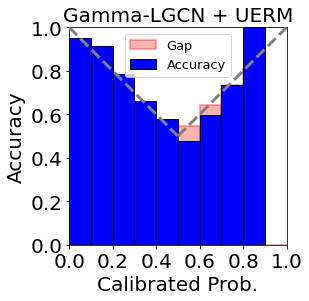}}
\caption{Reliability diagram of each calibration method. Gap denotes the discrepancy between the accuracy and the average calibrated probability for each bin. The grey dashed line is a diagonal function that indicates the ideal reliability line where the blue accuracy bar should meet.}
\end{figure*}
\section{Experiment}
\subsection{Experimental Setup}
We concisely introduce our experimental settings in this section.
For more details, please refer to Appendix E.
Our source code is publicly available\footnote{\url{https://github.com/WonbinKweon/CalibratedRankingModels_AAAI2022}}.
\subsubsection{Datasets}
To evaluate the calibration quality of predicted preference probability, we need an unbiased test set where we can directly observe the preference variable $R_{u,i}$ without any bias from the observation process $O_{u,i}$.
To the best of our knowledge, there are two real-world datasets that have separate unbiased test sets where the users are asked to rate uniformly sampled items (i.e., $O_{u,i}=1$ for test sets).
Note that in the training set, we only observe the interaction $Y_{u,i}$.
\textbf{Yahoo!R3}\footnote{\url{http://research.yahoo.com/Academic_Relations}} has over 300K \textit{interactions} in the training set and 54K \textit{preferences} in the test set from 15.4K users and 1K songs.
\textbf{Coat} \cite{ips16} has over 7K interactions in the training set and 4.6K preferences in the test set from 290 users and 300 coats.
We hold out 10\% of the training set as the validation set for the hyperparameter tuning of the base models and the optimization of the calibration methods.
\subsubsection{Base models}
For rigorous evaluation, we apply the calibration methods on several widely-used personalized ranking models with various model architectures and loss functions: Bayesian Personalized Ranking (BPR) \cite{bpr09}, Neural Collaborative Filtering (NCF) \cite{ncf17}, Collaborative Metric Learning (CML) \cite{cml17}, Unbiased BPR (UBPR) \cite{saito19}, and LightGCN (LGCN) \cite{lightgcn20}.
The details for the training of these base models can be found in Appendix E.
\subsubsection{Calibration methods compared}
We evaluate the proposed calibration methods with various calibration methods.
For the naive baseline, we adopt the minmax normalizer and the sigmoid function which simply re-scale the scores into [0,1] without calibration.
For non-parametric methods, we adopt Histogram binning \cite{hist01}, Isotonic regression \cite{iso12}, and BBQ \cite{bbq15}.
For parametric methods, we adopt Platt scaling \cite{platt99} and Beta calibration \cite{beta17}.
Note that we do not compare recent work designed for multi-class classification \cite{kull2019beyond, rahimi2020intra}, since they are either the generalized version of Beta calibration or cannot be directly adopted for the personalized ranking models.
\subsubsection{Evaluation metrics}
We adopt well-known calibration metrics like ECE, MCE with $M=15$, and NLL as done in recent work \cite{kull2019beyond, rahimi2020intra}.
We also plot the reliability diagram that shows the discrepancy between the accuracy and the average calibrated probability of each probability interval.
Note that evaluation metrics are computed on $R_{u,i}$ which is observed only from the test set.
\subsubsection{Evaluation process}
We first train the base personalized ranking model $f_{\theta}(u,i)$ with $Y_{u,i}$ on the training set.
Second, we compute ranking score $s_{u,i} = f_{\theta}(u,i)$ for user-item pairs in the validation set.
Third, we optimize the calibration method $g_{\phi}(s)$ on the validation set with the computed $s_{u,i}$ and the estimated $\hat{\omega}_{u,i}$, with $f_{\theta}(u,i)$ fixed.
Lastly, we evaluate the calibrated probability $p=g_{\phi}(s_{u,i})$ with $R_{u,i}$ from the unbiased test set by using the above evaluation metrics.

\subsection{Comparing Calibration Performance}
Table 1 shows ECE of each calibration method applied on the various personalized ranking models (MCE and NLL can be found in Appendix F).
ECE indicates how well the calibrated probabilities and ground-truth likelihoods match on the test set across all probability ranges.
First, the minmax normalizer and the sigmoid function produce poorly calibrated preference probabilities.
It is obvious because the ranking scores do not have any probabilistic meaning and naively re-scaling them cannot reflect the score distribution.

Second, the parametric methods better calibrate the preference probabilities than the non-parametric methods in most cases.
This is consistent with recent work \cite{cal17, kull2019beyond} for image classification.
The non-parametric calibration methods lack rich expressiveness since they rely on the binning scheme, which maps the ranking scores to the probabilities in a discrete manner.
On the other hand, the parametric calibration methods fit the continuous functions based on the parametric distributions.
Therefore, they have a more granular mapping from the ranking scores to the preference probabilities.

Third, every parametric calibration method benefits from adopting $\mathcal{L}_{\textup{UERM}}$ instead of $\mathcal{L}_{\textup{naive}}$ for the parameter fitting.
The naive log-loss treats all the unobserved pairs as negative pairs and makes the calibration methods produce biased preference probabilities.
On the contrary, inverse propensity-scored log-loss handles such problem and enables us to compute the ideal empirical risk indirectly.
As a result, ECE decreases by 7.40\%$\sim$76.52\% for all parametric methods compared to when the naive log-loss is used for the optimization.

Lastly, Gaussian calibration and Gamma calibration with $\mathcal{L}_{\textup{UERM}}$ show the best calibration performance across all base models and datasets.
Platt scaling can be seen as a special case of the proposed methods with $a=0$, so it has less expressiveness in terms of the capacity of parametric family.
Beta distribution is only defined in [0,1], so it cannot represent the unbounded ranking scores.
To adopt Beta calibration, we need to re-scale the ranking score, however, it is not verified for the optimality \cite{iso12}.
As a result, our calibration methods improve ECE by 5.21\%$\sim$25.85\% over the best competitor.
Also, since our proposed models have a larger capacity of expressiveness, they show larger improvement on Yahoo!R3, which has more samples to fit the parameters than Coat. 

\subsection{Reliability Diagram}
Figure 1 shows the reliability diagram \cite{cal17} for each calibration method applied on LGCN for Yahoo!R3.
We partition the calibrated probabilities $g_{\phi}(s)$ into 10 equi-spaced bins and compute the accuracy and the average calibrated probability for each bin (i.e., the first and the second term in Eq.4, respectively).
The accuracy is the same with the ground-truth proportion of positive samples for the positive bins (i.e., probability over 0.5) and the ground-truth proportion of negative samples for the negative bins (i.e., probability under 0.5).
Note that the bar does not exist if the bin does not have any prediction in it.

First, the non-parametric calibration methods do not produce the probability over 0.6.
It is because they can easily be overfitted to the unbalanced user-item interaction datasets since they do not have any prior distribution. 
On the other hand, the parametric calibration methods produce probabilities across all ranges by avoiding such overfitting problem with the prior parametric distributions.

Second, the parametric calibration methods with UERM produce well-calibrated probabilities especially for the positive preference ($p>0.5$).
The naive log-loss makes the calibration methods biased towards the negative preference, by treating all the unobserved pairs as the negative pairs.
As a result, the parametric methods with the naive log-loss (upper-right three diagrams of Figure 1) show large gaps in the positive probability range ($p>0.5$).
On the contrary, UERM framework successfully alleviates this problem and produces much smaller gaps for the positive preference (lower-right three diagrams of Figure 1).
Lastly, it is quite a natural result that parametric methods with UERM do not produce the probability over 0.9, considering that the users prefer only a few items among a large number of items.

\subsection{Score Distribution \& Fitted Function}
\begin{figure}[t]
\centering 
\subfigure{\includegraphics[width=0.4\linewidth]{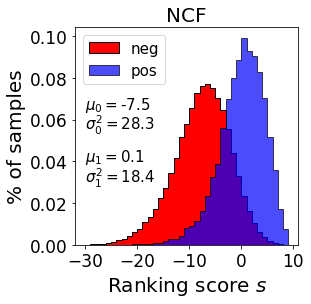}}
\subfigure{\includegraphics[width=0.4\linewidth]{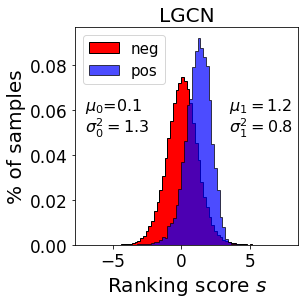}}
\caption{Ranking score distributions of negative and positive pairs.}
\end{figure}

\begin{figure}[t]
\centering 
\subfigure{\includegraphics[width=0.3\linewidth]{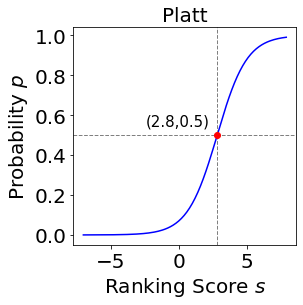}}
\subfigure{\includegraphics[width=0.3\linewidth]{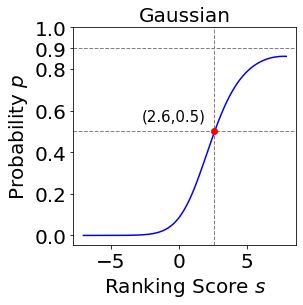}}
\subfigure{\includegraphics[width=0.3\linewidth]{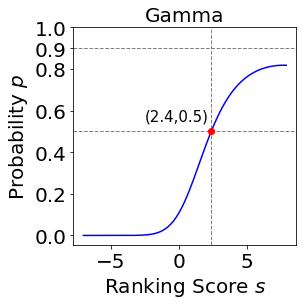}}
\caption{Fitted function of each calibration method.}
\end{figure}
Figure 2 shows the distribution of ranking scores trained by NCF and LGCN on Yahoo!R3.
We can see that the class-conditional score distributions have different deviations ($\sigma_0 > \sigma_1$) and skewed shapes (left tails are longer than the right tails).
This indicates that Platt scaling (or temperature scaling) assuming the same variance for both classes cannot effectively handle these score distributions.
Figure 3 shows the fitted calibration function of each parametric method adopted on LGCN and optimized with UERM.
Since most of the user-item pairs are negative in the interaction datasets, all three functions are fitted to produce the low probability under 0.1 for a wide bottom range to reflect the dominant negative preferences.
Platt scaling is forced to have the symmetric shape due to its parametric family, so it produces the high probability over 0.9 which is symmetrical to that of under 0.1.
On the other hand, Gaussian calibration and Gamma calibration, which have a larger expressive power, learn asymmetric shapes tailored to the score distributions having different deviations and skewness.
This result shows that they effectively handle the imbalance of user-item interaction datasets and supports the experimental superiority of the proposed methods.

\subsection{Case Study}
\begin{figure}[t]
\centering 
\includegraphics[width=0.65\linewidth]{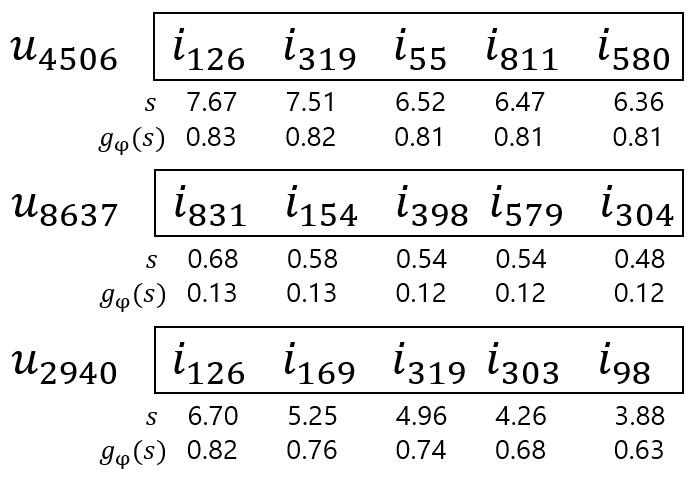}
\caption{Case study. Top-5 items for each user with ranking score $s$ and calibrated probability $g_{\phi}(s)$.}
\end{figure}
Figure 4 shows the case study on Yahoo!R3 with Gaussian calibration adopted on LGCN.
The personalized ranking model first learns the ranking scores and produces a top-5 ranking list for each user.
Then, Gaussian calibration transforms the ranking scores to the well-calibrated preference probabilities.
For the first user $u_{4506}$, the method produces high preference probabilities for all top-5 items.
In this case, we can recommend them to him with confidence.
On the other hand, for the second user $u_{8637}$, all top-5 items have low preference probabilities, and the last user $u_{2940}$ has a wide range of preference probabilities.
For these users, merely recommending all the top-ranked items without consideration of potential preference degrade their satisfaction.
It is also known that the unsatisfactory recommendations even make the users leave the platform \cite{leave06}.
Therefore, instead of recommending items with low confidence, the system should take other strategies, such as requesting additional user feedback \cite{dre}.

\section{Conclusion}
In this paper, we aim to obtain calibrated probabilities with personalized ranking models.
We investigate various parametric distributions and propose two parametric calibration methods, namely Gaussian calibration and Gamma calibration. 
We also design the unbiased empirical risk minimization framework that helps the calibration methods to be optimized towards true preference probability with the biased user-item interaction dataset.
Our extensive evaluation demonstrates that the proposed methods and framework significantly improve calibration metrics and have a richer expressiveness than existing methods.
Lastly, our case study shows that the calibrated probability provides an objective criterion for the reliability of recommendations, allowing the system to take various strategies to increase user satisfaction.

\section{Acknowledgments}
This work was supported by the NRF grant funded by the MSIT (South Korea, No.2020R1A2B5B03097210), the IITP grant funded by the MSIT (South Korea, No.2018-0-00584, 2019-0-01906), and the Technology Innovation Program funded by the MOTIE (South Korea, No.20014926).

\bibliography{aaai22}

\clearpage
\appendix
\section{Appendix}
\subsection{A. Satisfaction of Proposed Desiderata}
\begin{table}[h]
    \centering\fontsize{9}{10}\selectfont
    \begin{tabular}{cccc}
        \toprule
        \multirow{2}{*}{Method} & (1) & (2) & (3) \\ 
         & \multirow{1}{*}{Input range} & \multirow{1}{*}{Monotonicity} & \multirow{1}{*}{Expressiveness} \\ 
        \midrule
        Hist & \checkmark & & $\triangle$ \\
        Isotonic & \checkmark & \checkmark & $\triangle$\\
        BBQ & & & $\triangle$ \\
        \midrule
        Platt & \checkmark & \checkmark & $\triangle$\\
        Temp. S & \checkmark & \checkmark & $\triangle$ \\
        Beta & & \checkmark & \checkmark\\
        \midrule
        Gaussian & \checkmark & \checkmark & \checkmark \\
        Gamma & \checkmark & \checkmark & \checkmark \\
        \bottomrule
    \end{tabular}
    \caption{Satisfaction of proposed desiderata for each calibration method. $\checkmark$ indicates the satisfaction, $\triangle$ means the insufficiency, and a blank denotes the unsatisfaction.}
\end{table}
\noindent The non-parametric calibration methods (Hist, Isotonic, and BBQ) lack rich expressiveness since they rely on the binning scheme, which maps the ranking scores to the probabilities in a discrete manner.
Also, histogram binning \cite{hist01} and BBQ \cite{bbq15} do not guarantee the monotonicity and BBQ takes the input only from [0,1].

Platt scaling \cite{platt99} and temperature scaling \cite{cal17} assume Gaussian distributions with the same variance for positive and negative classes, therefore, they do not have enough expressiveness for modeling the different variances and the skewness of class-conditional distributions.
Beta calibration takes the input only from [0,1] so it cannot handle the unbounded ranking scores.
Our proposed methods (Gaussian and Gamma) learn the asymmetric shape that handles the imbalance in interaction datasets and satisfy all the desiderata.

\subsection{B. Other Distributions}
Besides Gaussian and Gamma distribution, Swets \cite{exp69} adopts Exponential distributions for the score distribution of both classes:
\begin{equation}
\begin{split}
    & p(s|Y=0) = \lambda_0 \text{exp}(-\lambda_0 s), \\
    & p(s|Y=1) = \lambda_1 \text{exp}(-\lambda_1 s), \\
\end{split}
\end{equation}
where $\lambda_0, \lambda_1 \in \mathbb{R}^+$ are the rate parameters for each Exponential distribution.
Note that for Gamma distribution and Exponential distribution, we need to shift the score to make all the inputs positive: $s \leftarrow s-s_{min}$, where $s_{min}$ is the minimum ranking score. 
Then, the posterior can be computed as follows:
\begin{equation}
\begin{split}
    P(Y=1|s) & = \frac{1}{1 + \pi_0 p(s|Y=0) / \pi_1 p(s|Y=1)} \\
    & = \frac{1}{1 + \text{exp}[ (\lambda_1 - \lambda_0)s + \text{log}(\pi_0 \lambda_0 / \pi_1 \lambda_1) ]} \\
    & = \sigma(bs + c),
\end{split}
\end{equation}
where $b=\lambda_0 - \lambda_1$ and $c=\text{log}(\pi_1 \lambda_1 / \pi_0 \lambda_0) \in \mathbb{R}$.
It is the exact same form as Platt scaling.

On the other hand, Manmatha \cite{normexp01} proposes Exponential distribution for the negative class and Gaussian distribution for the positive class:
\begin{equation}
\begin{split}
    & p(s|Y=0) = \lambda_0 \text{exp}(-\lambda_0 s), \\
    & p(s|Y=1) = (\sqrt{2\pi}\sigma_1)^{-1} \text{exp} [ - (s-\mu_1)^2 / 2\sigma_1^2 ],\\
\end{split}
\end{equation}
where $\lambda_0, \mu_1, \sigma_1^2 \in \mathbb{R}^+$.
Then, we have the posterior as follows:
\begin{equation}
\begin{split}
    P(Y=1|s) & = \frac{1}{1 + \pi_0 p(s|Y=0) / \pi_1 p(s|Y=1)} \\
    & = \frac{1}{1 + \frac{\pi_0 \lambda_0 \sqrt{2\pi} \sigma_1}{\pi_1} \text{exp}[(s-\mu_1)^2 / 2\sigma_1^2 - \lambda_0 s] } \\
    & = \frac{1}{1 + \text{exp}[ (2\sigma_1^2)^{-1}s^2 - (\lambda_0 + \mu_1/\sigma_1^2)s - c ]} \\ 
    & = \sigma(as^2 + bs + c),
\end{split}
\end{equation}
where $a=(-2\sigma_1^2)^{-1} \in \mathbb{R}^-$, $b=\lambda_0 + \mu_1/\sigma_1^2 \in \mathbb{R}^+$, and $c=\text{log}(\pi_1 / \pi_0 \lambda_0 \sqrt{2\pi} \sigma_1) - \mu_1^2/(2\sigma_1^2)\in \mathbb{R}$.
With the constraints $a<0$ and $b>0$, the calibration function in Eq.4 is cannot be monotonically increasing on $s>0$ with any parameters.

Lastly, Kanoulas \cite{gammagaussian10} proposes gamma distribution for the negative class and Gaussian distribution for the positive class:
\begin{equation}
\begin{split}
    & p(s|Y=0) = \rho(\alpha_0)^{-1} \beta_0^{\alpha_0} s^{\alpha_0-1} \text{exp}(-\beta_0s), \\
    & p(s|Y=1) = (\sqrt{2\pi}\sigma_1)^{-1} \text{exp} [ - (s-\mu_1)^2 / 2\sigma_1^2 ],\\
\end{split}
\end{equation}
where $\alpha_0$, $\beta_0$, $\mu_1$, $\sigma_1^2 \in \mathbb{R}^{+}$.
The posterior for these likelihoods can be computed as follows:
\begin{equation}
\begin{split}
    P(Y=1|s) & = \frac{1}{1 + \pi_0 p(s|Y=0) / \pi_1 p(s|Y=1)} \\
    & = \frac{1}{1 + \text{exp}[ \frac{1}{2\sigma_1^2}s^2 - (\beta_0 + \frac{\mu_1}{\sigma_1^2})s + (\alpha_0 -1)\text{log}s - c]} \\
    & = \sigma(as^2 + bs + b'\text{log}s + c),
\end{split}
\end{equation}
where $a=(-2\sigma_1^2)^{-1} \in \mathbb{R}^-$, $b=\beta_0+\mu_1/\sigma_1^2 \in \mathbb{R}^+$, $b'=1-\alpha_0 \in \mathbb{R}$, and $c=\text{log}(\pi_1 \rho(\alpha_0) / \pi_0\beta_0^{\alpha_0} \sigma_1 \sqrt{2\pi}) - \mu_1^2/(2\sigma_1^2) \in \mathbb{R}$.
For this function to be monotonically increasing, we need a non-linear constraint $b'- \frac{b^2}{8a} < 0$ (derivation of this constraint can be easily done by taking the derivative of Eq.22 w.r.t. the score $s$).
Since the optimization of logistic regression with non-linear constraints is not straightforward, it is hard for this posterior to satisfy the second desideratum.

\subsection{C. Monotonicity for Proposed Desiderata}
Gaussian calibration $g_{\phi}(s) = \sigma(as^2+bs+c)$ is monotonically increasing if and only if the parameter $a$ and $b$ satisfy the constraint $2as+b > 0$ for $s_{\textup{min}}$ and $s_{\textup{max}}$.
\begin{proof}
\begin{equation*}
\begin{split}
    & g_{\phi}'(s) = \sigma'(as^2+bs+c) \cdot (2as+b) > 0 \\
    & \Longleftrightarrow 2as+b > 0 \\
    & \because \; \sigma'(x) = \sigma(x)(1-\sigma(x)) > 0 \;\; \textup{for all} \; x \in \mathbb{R} \\
    & \Longleftrightarrow 2as_{\textup{min}}+b > 0 \textup{ and } 2as_{\textup{max}}+b > 0 \\
    & \because 2as + b \textup{ is a linear function of s in } [s_{\textup{min}}, s_{\textup{max}}]. \\
\end{split}
\end{equation*}
\end{proof}
Gamma calibration $g_{\phi}(s) = \sigma(a\textup{log}s+bs+c)$ is monotonically increasing if and only if the parameter $a$ and $b$ satisfy the constraint $a/s+b > 0$ for $s_{\textup{min}}$ and $s_{\textup{max}}$.
\begin{proof}
\begin{equation*}
\begin{split}
    & g_{\phi}'(s) = \sigma'(a\textup{log}s+bs+c) \cdot (a/s+b) > 0 \\
    & \Longleftrightarrow a/s+b > 0 \\
    & \because \; \sigma'(x) = \sigma(x)(1-\sigma(x)) > 0 \;\; \textup{for all} \; x \in \mathbb{R} \\
    & \Longleftrightarrow a/s_{\textup{min}}+b > 0 \textup{ and } a/s_{\textup{max}}+b > 0 \\
    & \because a/s + b \textup{ is a monotonic function of s in } [s_{\textup{min}}, s_{\textup{max}}]. \\
\end{split}
\end{equation*}
\end{proof}

\subsection{D. Unbiased Empirical Risk Minimization}
\setcounter{proposition}{0}
\begin{proposition}
$\hat{\mathcal{R}}_{\textup{UERM}}(g_{\phi}|\omega)$, which is the empirical risk of $\mathcal{L}_{\textup{UERM}}$ on validation set with true propensity score $\omega$, is equal to $\hat{\mathcal{R}}_{\textup{ideal}}(g_{\phi})$, which is the ideal empirical risk.
\end{proposition}
\begin{proof}
\begin{equation*}
\begin{split}
    \hat{\mathcal{R}}_{\text{UERM}}(g_{\phi}|\omega) = & \: \mathbb{E}_{(u,i) \in \mathcal{D}_{\text{val}}} \big[\mathcal{L}_{\text{IPS}}(u,i)\big] \\
    = & - \frac{1}{\abs{\mathcal{D}_{\text{val}}}} \sum_{(u,i) \in \mathcal{D}_{\text{val}}} \mathbb{E} \left[ \frac{Y_{u,i}}{\omega_{u,i}} \text{log}( g_{\phi}(s_{u,i})) \right.\\
    & + \left. (1-\frac{Y_{u,i}}{\omega_{u,i}}) \text{log}(1 - g_{\phi}(s_{u,i})) \right]\\
    = & - \frac{1}{\abs{\mathcal{D}_{\text{val}}}} \sum_{(u,i) \in \mathcal{D}_{\text{val}}} \frac{\omega_{u,i}\rho_{u,i}}{\omega_{u,i}} \text{log}( g_{\phi}(s_{u,i})).\\
    & + (1-\frac{\omega_{u,i}\rho_{u,i}}{\omega_{u,i}}) \text{log}(1 - g_{\phi}(s_{u,i})) \\
    = & - \frac{1}{\abs{\mathcal{D}_{\text{val}}}} \sum_{(u,i) \in \mathcal{D}_{\text{val}}} \rho_{u,i} \text{log}( g_{\phi}(s_{u,i})) \\
    & + (1-\rho_{u,i}) \text{log}(1 - g_{\phi}(s_{u,i})) \\
    = & \: \mathbb{E}_{(u,i) \in \mathcal{D}_{\text{val}}} \big[\mathcal{L}_{\text{ideal}}(u,i)\big] \\
    = & \: \hat{\mathcal{R}}_{\text{ideal}}(g_{\phi}).
\end{split}
\end{equation*}
\end{proof}
\begin{proposition}
The bias of $\, \hat{\mathcal{R}}_{\textup{UERM}}(g_{\phi}|\hat{\omega})$ induced by the inaccurately estimated propensity scores $\hat{\omega}$ is  $\frac{1}{\abs{\mathcal{D}_{\textup{val}}}} \sum_{(u,i) \in \mathcal{D}_{\textup{val}}} \rho_{u,i} \left(\frac{\omega_{u,i}}{\hat{\omega}_{u,i}}-1 \right) \textup{log}\left( \frac{g_{\phi}(s_{u,i})}{1-g_{\phi}(s_{u,i})} \right)$.
\end{proposition}
\begin{proof}
\begin{equation*}
\begin{split}
    \text{bias} = & \, \hat{\mathcal{R}}_{\text{UERM}}(g_{\phi}|\hat{\omega}) - \hat{\mathcal{R}}_{\text{ideal}}(g_{\phi}) \\
    = & \, \frac{1}{\abs{\mathcal{D}_{\text{val}}}} \sum_{(u,i) \in \mathcal{D}_{\text{val}}} - \frac{\omega_{u,i}\rho_{u,i}}{\hat{\omega}_{u,i}} \text{log}(g_{\phi}(s_{u,i})) \\
    & - (1-\frac{\omega_{u,i}\rho_{u,i}}{\hat{\omega}_{u,i}})\text{log}(1-g_{\phi}(s_{u,i})) + \rho_{u,i}\text{log}(g_{\phi}(s_{u,i})) \\
    & + (1-\rho_{u,i})\text{log}(1-g_{\phi}(s_{u,i})) \\
    = & \, \frac{1}{\abs{\mathcal{D}_{\text{val}}}} \sum_{(u,i) \in \mathcal{D}_{\text{val}}} (\omega_{u,i}/\hat{\omega}_{u,i}-1) [-\rho_{u,i}\text{log}(g_{\phi}(s_{u,i})) \\
    & + \rho_{u,i}\text{log}(1-g_{\phi}(s_{u,i}))] \\
    = & \, \frac{1}{\abs{\mathcal{D}_{\text{val}}}} \sum_{(u,i) \in \mathcal{D}_{\text{val}}} \rho_{u,i} \left(\frac{\omega_{u,i}}{\hat{\omega}_{u,i}}-1 \right) \text{log}\left( \frac{1-g_{\phi}(s_{u,i})}{g_{\phi}(s_{u,i})} \right) \\
\end{split}
\end{equation*}
\end{proof}

\subsection{E. Experimental Setup}
\subsubsection{Datasets}
For the training and the test set, we convert the ratings over 3 to $Y=1$, and the ratings under 4 to $Y=0$ as done in the conventional papers \cite{ncf17, saito19}.
\subsubsection{Base models}
BPR \cite{bpr09} learns the user and the item embeddings with BPR loss:
\begin{equation}
    \mathcal{L}_{\textup{BPR}} = \sum_{u \in \mathcal{U}, i,j \in \mathcal{I}} - \text{log} \sigma(f_\theta(u,i) - f_\theta(u,j)) Y_{u,i} (1-Y_{u,j}),
\end{equation}
where $f_\theta(u,i) = \textbf{e}_u \cdot \textbf{e}_i$ and $\textbf{e}_u$, $\textbf{e}_i$ are user and item embeddings.

NCF \cite{ncf17} learns the ranking score of a user-item pair with the binary cross-entropy loss:
\begin{equation}
\begin{split}
      \mathcal{L}_{\textup{NCF}} & = \sum_{u \in \mathcal{U}, i \in \mathcal{I}} -Y_{u,i}\textup{log}\sigma(f_\theta(u,i)) \\ & -(1-Y_{u,i})\textup{log}(1-\sigma(f_\theta(u,i))),  
\end{split}
\end{equation}
where $f_\theta(u,i) = \textup{NeuralNet}(\textbf{e}_u, \textbf{e}_i)$.

CML \cite{cml17} learns the user and the item embeddings with the triplet loss:
\begin{equation}
    \mathcal{L}_{\textup{CML}} = \sum_{u \in \mathcal{U}, i,j \in \mathcal{I}} \textup{max}(0, m+f_\theta(u,i) - f_\theta(u,j)) Y_{u,i} (1-Y_{u,j}),
\end{equation}
where $m$ is the margin and $f_\theta(u,i) = ||\textbf{e}_u - \textbf{e}_i||^2$.

UBPR \cite{saito19} learns the user and the item embeddings with the inverse propensity-scored BPR loss:
\begin{equation}
    \mathcal{L}_{\textup{UBPR}} = \sum_{u \in \mathcal{U}, i,j \in \mathcal{I}} - \text{log} \sigma(f_\theta(u,i) - f_\theta(u,j)) \frac{Y_{u,i}}{\omega_{u,i}} (1-\frac{Y_{u,i}}{\omega_{u,i}}),
\end{equation}
where $f_\theta(u,i) = \textbf{e}_u \cdot \textbf{e}_i$ and $\omega_{u,i}$ is the propensity score estimated by the same technique used in our framework.

LGCN \cite{lightgcn20} learns the user and the item embeddings with BPR loss:
\begin{equation}
    \mathcal{L}_{\textup{LGCN}} = \sum_{u \in \mathcal{U}, i,j \in \mathcal{I}} - \text{log} \sigma(f_\theta(u,i) - f_\theta(u,j)) Y_{u,i} (1-Y_{u,j}),
\end{equation}
where $f_\theta(u,i) = \textup{LGCN}(\textbf{e}_u) \cdot \textup{LGCN}(\textbf{e}_i)$ and  $\textup{LGCN}(\cdot)$ is simplified Graph Convolutional Networks (GCN).

\begin{table*}[t!]
    \centering\fontsize{9}{10}\selectfont
    \begin{tabular}{c|ccccc|ccccc}
        \toprule
         & \multicolumn{5}{c|}{Yahoo!R3} & \multicolumn{5}{c}{Coat} \\
        \toprule
        Metric & BPR & NCF & CML & UBPR & LGCN & BPR & NCF & CML & UBPR & LGCN \\
        \midrule
        NDCG@1 & 0.5070 & 0.5181 & 0.5259 & 0.5328 & \textbf{0.5443} & 0.3924 & 0.5341 & \textbf{0.5865} & 0.3982 & 0.4008 \\
        NDCG@3 & 0.5519 & 0.5812 & 0.5716 & 0.5919 & \textbf{0.6002} & 0.3761 & \textbf{0.5129} & 0.5020 & 0.3962 & 0.3973 \\
        NDCG@5 & 0.6176 & 0.6467 & 0.6379 & 0.6555 & \textbf{0.6637} & 0.4302 & \textbf{0.5452} & 0.5142 & 0.4418 & 0.4301 \\
        Recall@1 & 0.3126 & 0.3213 & 0.3286 & 0.3345 & \textbf{0.3395} & 0.1321 & 0.2132 & \textbf{0.2334} & 0.1354 & 0.1395 \\ 
        Recall@3 & 0.5743 & 0.6098 & 0.5918 & 0.6207 & \textbf{0.6280} & 0.2852 & \textbf{0.4166} & 0.3847 & 0.3181 & 0.3113 \\ 
        Recall@5 & 0.7428 & 0.7779 & 0.7613 & 0.7820 & \textbf{0.7915} & 0.4638 & \textbf{0.5146} & 0.4847 & 0.4757 & 0.4816 \\
        \bottomrule
    \end{tabular}
    \caption{Ranking performance of each personalized ranking model. Numbers in boldface are the best results.}
\end{table*}

For the base models, we basically follow the source code of the authors.
NCF has 2-layer MLP for MLP module and 1-layer MLP for the prediction module.
LGCN has a 2-layer simplified GCN for the training and the inference.
We use 128 for the size of user and item embeddings for all base models, except NCF which adopts 64 for the embedding size.
The batch size is 512, the learning rate is 0.001, the weight decay rate is 0.001, and the negative sample rate is 1.
Each model is trained until the convergence and their ranking performance can be found in Table 3.
LGCN shows the best performance for Yahoo!R3 dataset and NCF or CML shows the best performance for Coat dataset.

\subsubsection{Calibration method compared}
For Histogram binning, we set the number of bins as 50 for Yahoo!R3, 15 for Coat.
For BBQ, we set the number of binning methods as 4 and the number of bins of each binning method is (10,20,50,100).
For those which do not meet the first desideratum (BBQ and Beta calibration), we adopt the sigmoid function to re-scale the input into [0,1].

\subsubsection{Computing  infrastructures}
We adopt a Titan X GPU and an Intel(R) Core(TM) i7-7820X 3.60GHz CPU.
Optimization of all calibration methods is done in at most a few minutes.

\begin{table*}[h!]
    \centering\fontsize{9}{10}\selectfont
    \begin{tabular}{cc|ccccc|ccccc}
        \toprule
         & & \multicolumn{5}{c|}{Yahoo!R3} & \multicolumn{5}{c}{Coat} \\
        \toprule
        Type & Methods & BPR & NCF & CML & UBPR & LGCN & BPR & NCF & CML & UBPR & LGCN \\
        \midrule
        \multirow{2}{*}{uncalibrated} & MinMax & 0.4929 & 0.4190 & 0.3152 & 0.3004 & 0.2258 & 0.1790 & 0.4624 & 0.1834 & 0.1920 & 0.2350\\
         & Sigmoid & 0.5726 & 0.5248 & 0.2571 & 0.7103 & 0.5082 & 0.3868 & 0.6223 & 0.6707 & 0.2699 & 0.4642\\
        \midrule
        \multirow{3}{*}{non-parametric} & Hist & 0.2620 & 0.2369 & 0.6184 & 0.1349 & 0.1366 & 0.2492 & 0.4083 & 0.4924 & 0.2910 & 0.3940\\
         & Isotonic & 0.2136 & 0.2173 & 0.5252 & 0.1909 & 0.1270 & 0.2495 & 0.4246 & 0.4812 & 0.2547 & 0.3500\\
         & BBQ & 0.2116 & 0.2489 & 0.7076 & 0.1701 & 0.1157 & 0.2545 & 0.3749 & 0.3423 & 0.3358 & 0.2589\\
        \midrule
        \multirow{4}{*}{\shortstack{parametric \\ w/ $\mathcal{L}_{\textup{naive}}$}} & Platt & 0.2032 & 0.2876 & 0.3254 & 0.3117 & 0.3068 & 0.3664 & 0.1844 & 0.5796 & 0.4079 & 0.4633\\
         & Beta & 0.2563 & 0.2575 & 0.5049 & 0.2705 & 0.3033 & 0.4071 & 0.2175 & 0.5782 & 0.4352 & 0.3367\\
         & Gaussian & 0.2387 & 0.2366 & 0.3737 & 0.3282 & 0.2200 & 0.3177 & 0.1678 & 0.5878 & 0.4807 & 0.3854\\
         & Gamma & 0.2123 & 0.2260 & 0.3149 & 0.3126 & 0.2962 & 0.3616 & 0.1702 & 0.5968 & 0.4185 & 0.3480\\
        \midrule
          & Platt & 0.1951 & 0.2504 & 0.2293 & 0.1257 & 0.1143 & 0.2625 & 0.1882 & 0.4423 & 0.2708 & 0.2225\\
        parametric & Beta & 0.2004 & 0.2453 & 0.2734 & 0.1317 & 0.1905 & 0.3278 & 0.2085 & 0.4150 & 0.2574 & 0.2501\\
        w/ $\mathcal{L}_{\textup{UERM}}$ & \cellcolor{gray!20}Gaussian & \cellcolor{gray!20}0.1816 & \cellcolor{gray!20}0.2259 & \cellcolor{gray!20}0.2451 & \cellcolor{gray!20}\textbf{0.1231} & \cellcolor{gray!20}\textbf{0.1064} & \cellcolor{gray!20}\textbf{0.2380} & \cellcolor{gray!20}\textbf{0.1236} & \cellcolor{gray!20}\textbf{0.3390} & \cellcolor{gray!20}\textbf{0.2419} & \cellcolor{gray!20}0.2117\\
         & \cellcolor{gray!20}Gamma & \cellcolor{gray!20}\textbf{0.1592} & \cellcolor{gray!20}\textbf{0.2074} & \cellcolor{gray!20}\textbf{0.2268} & \cellcolor{gray!20}\textbf{0.1231} &\cellcolor{gray!20} 0.1118 & \cellcolor{gray!20}0.2543 & \cellcolor{gray!20}\textbf{0.1236} &\cellcolor{gray!20}0.4663 &\cellcolor{gray!20}0.2478 & \cellcolor{gray!20}\textbf{0.2023}\\
        \bottomrule
    \end{tabular}
    \caption{Maximum Calibration Error with $M=15$ of each calibration method applied on five personalized ranking models. Numbers in boldface are the best results.}
\end{table*}
\begin{table*}[h!]
    \centering\fontsize{9}{10}\selectfont
    \begin{tabular}{cc|ccccc|ccccc}
        \toprule
         & & \multicolumn{5}{c|}{Yahoo!R3} & \multicolumn{5}{c}{Coat} \\
        \toprule
        Type & Methods & BPR & NCF & CML & UBPR & LightGCN & BPR & NCF & CML & UBPR & LightGCN \\
        \midrule
        \multirow{2}{*}{uncalibrated} & MinMax & 0.4929 & 0.4190 & 0.3152 & 0.3004 & 0.2258 & 0.1790 & 0.4624 & 0.1834 & 0.1920 & 0.2350\\
         & Sigmoid & 0.7030 & 0.4641 & 0.3072 & 0.7083 & 0.6890 & 0.6727 & 1.0978 & 0.4819 & 0.6780 & 0.6904\\
        \midrule
        \multirow{3}{*}{non-parametric} & Hist & 0.2753 & 0.2717 & 0.3361 & 0.2728 & 0.2769 & 0.6264 & 0.5116 & 0.5412 & 0.6177 & 0.5077\\
         & Isotonic & 0.2726 & 0.2723 & 0.3240 & 0.2728 & 0.2683 & 0.5110 & 0.4953 & 0.4711 & 0.5224 & 0.5386\\
         & BBQ & 0.2725 & 0.2693 & 0.3517 & 0.2749 & 0.2691 & 0.4867 & 0.4790 & 0.4895 & 0.5092 & 0.4813\\
        \midrule
        \multirow{4}{*}{\shortstack{parametric \\ w/ $\mathcal{L}_{\textup{naive}}$}} & Platt & 0.2756 & 0.2700 & 0.3184 & 0.2735 & 0.2671 & 0.4748 & 0.4771 & 0.4747 & 0.4735 & 0.4758\\
         & Beta & 0.2755 & 0.2697 & 0.3250 & 0.2738 & 0.2673 & 0.4766 & 0.4796 & 0.4747 & 0.4741 & 0.4776\\
         & Gaussian & 0.2748 & 0.2676 & 0.3196 & 0.2725 & 0.2671 & 0.4766 & 0.4768 & 0.4755 & 0.4730 & 0.4761\\
         & Gamma & 0.2735 & 0.2699 & 0.3181 & 0.2735 & 0.2672 & 0.4754 & 0.4764 & 0.4739 & 0.4747 & 0.4749\\
        \midrule
          & Platt & 0.2749 & 0.2684 & 0.3034 & 0.2723 & 0.2675 & 0.4746 & 0.4655 & 0.4654 & 0.4716 & 0.4741\\
        parametric & Beta & 0.2747 & 0.2685 & 0.3022 & 0.2721 & 0.2672 & 0.4744 & 0.4684 & 0.4665 & 0.4713 & 0.4751\\
        w/ $\mathcal{L}_{\textup{UERM}}$ & \cellcolor{gray!20}Gaussian & \cellcolor{gray!20}0.2743 & \cellcolor{gray!20}\textbf{0.2666} & \cellcolor{gray!20}\textbf{0.3021} & \cellcolor{gray!20}\textbf{0.2715} & \cellcolor{gray!20}\textbf{0.2671} & \cellcolor{gray!20}\textbf{0.4735} & \cellcolor{gray!20}\textbf{0.4619} & \cellcolor{gray!20}0.4653 & \cellcolor{gray!20}\textbf{0.4711} & \cellcolor{gray!20}\textbf{0.4727}\\
         & \cellcolor{gray!20}Gamma & \cellcolor{gray!20}\textbf{0.2723} & \cellcolor{gray!20}0.2681 & \cellcolor{gray!20}0.3035 & \cellcolor{gray!20}0.2720 &\cellcolor{gray!20}0.2674 & \cellcolor{gray!20}0.4742& \cellcolor{gray!20}0.4649 &\cellcolor{gray!20}\textbf{0.4651} &\cellcolor{gray!20}\textbf{0.4711} & \cellcolor{gray!20}0.4733\\
        \bottomrule
    \end{tabular}
    \caption{Negative Log-Likelihood of each calibration method applied on five personalized ranking models. Numbers in boldface are the best results.}
\end{table*}

\subsection{F. Additional Experimental Result}
\subsubsection{MCE and NLL of Each Calibration Method}
In Tables 4 and 5, we report MCE and NLL of each calibration method adopted on various personalized ranking models for two real-world datasets.
Similar to ECE, our proposed calibration methods (Gaussian and Gamma calibration) also show the best results in MCE and NLL.

\subsubsection{Propensity Estimation}
\begin{figure}[t]
\centering 
\includegraphics[width=0.9\linewidth]{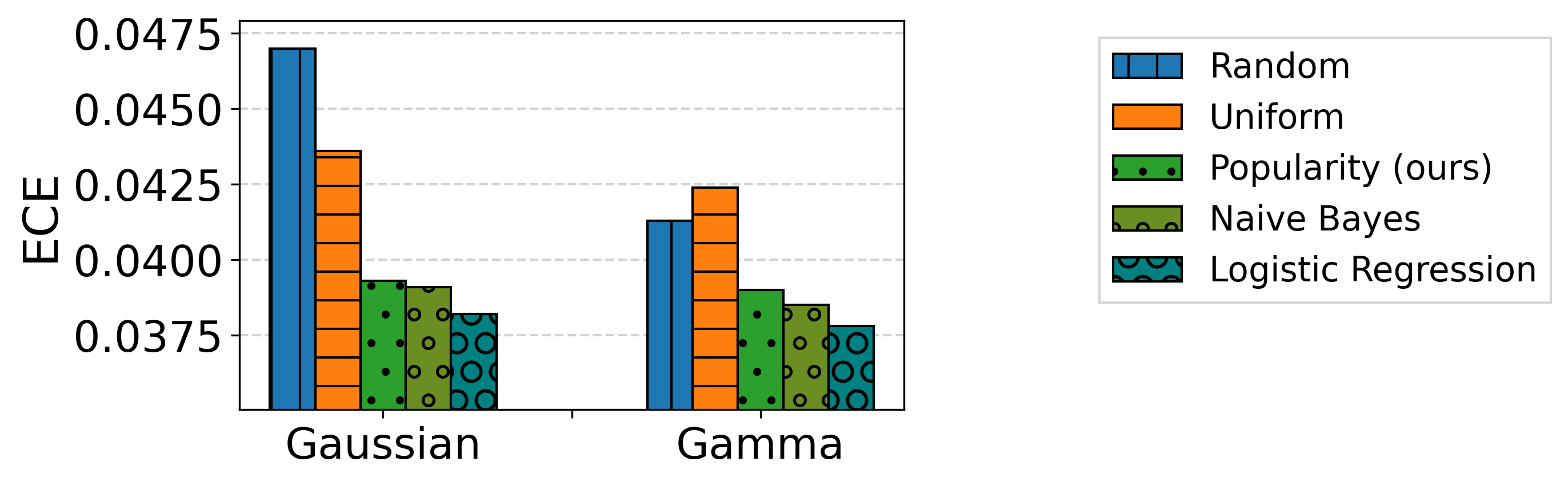}
\caption{ECE with various propensity estimation techniques. In this work, we utilize item popularities to estimate the propensity scores.}
\end{figure}
Figure 5 shows the ECE of the proposed methods adopted on BPR for Coat dataset with various propensity estimation techniques.
Random estimation produces random propensity scores from [0,1].
Uniform denotes $\omega=1$, which is equivalent to the naive log-loss.
In this work, we utilize item popularities to estimate the propensity scores.
Naive Bayes exploits the test set to compute the conditional probabilities.
Lastly, Logistic Regression uses user demographics and item categories to estimate the propensity scores.
Note that Naive Bayes and Logistic Regression utilize the additional information which is not available in our setting, and their propensity scores are provided by the author \cite{ips16}.
Surprisingly, merely utilizing popularity is enough for the estimation and it shows the comparable performance with Naive Bayes and Logistic Regression which use additional information.

\end{document}